# Interpretable AI predicts a 2026 summer dry anomaly in central China

Anran WANG [1†], Wen SHI [1†], Yong LUO [1*], Jianbin HUANG [2,3], Lijuan CHEN [4,5], Junhu ZHAO [5], Weixin JIN [6], and Huihui YUAN [1]

1 *Department of Earth System Science, Tsinghua University, Beijing* 100084*, China*

2 *Beijing Yanshan Earth Critical Zone National Research Station, University of Chinese Academy of Sciences, Beijing* 101408*, China*

3 *College of Resources and Environment, University of Chinese Academy of Sciences, Beijing* 101408*, China*

4 *State Key Laboratory of Climate System Prediction and Risk Management, National Climate Centre, China Meteorological Administration, Beijing* 100081*, China*

5 *China Meteorological Administration Key Laboratory for Climate Prediction Studies, National Climate Centre, China Meteorological Administration, Beijing* 100081*, China*

6 *Microsoft, Beijing* 100080*, China*

***Abstract***

Seasonal precipitation anomalies are largely regulated by atmospheric circulation, which dynamical models predict with greater reliability than precipitation itself. Here, we employ a deep learning model that translates dynamical circulation predictions into precipitation estimates. Predictions initialized from March to May consistently indicate a dry anomaly over central China in summer 2026. Retrospective evaluations revealed higher predictive skill in the analogue years, which also tended to feature central equatorial Pacific warming persisting from the preceding winter into summer. This warming favors an anomalous cyclonic circulation over the western North Pacific–South China Sea–South China region, which induces northerly winds and moisture divergence that jointly suppress rainfall over central China. Supporting this mechanism, layer-wise relevance propagation (LRP) independently identifies these northerly winds as the dominant driver of the prediction among all model inputs. Perturbation tests supported this attribution: removing LRP-identified features effectively eliminates the dry anomaly. Our framework thus provides physically interpretable explanations for AI-derived regional climate projections, facilitating evidence-based assessment before observational data become available.

[†] These authors contributed equally: Anran Wang and Wen Shi.

[*] Correspondence to: Yong Luo (yongluo@tsinghua.edu.cn)

## *Introduction*

Credible and skillful precipitation predictions issued ahead of the summer flood season in China could provide actionable lead time for water management, agricultural planning, and drought and flood preparedness[1,2]. Yet regional summer precipitation remains one of the most challenging targets of seasonal prediction owing to complex interactions among atmospheric, oceanic and land-surface processes[3–6]. Seasonal precipitation skill is generally modest and varies substantially with the prevailing climate state[7], creating intermittent windows of opportunity for more skillful prediction[8–11]. Averaging skill over a hindcast period can obscure this state-dependent predictability and provide only limited guidance on the credibility of an individual real-time prediction[12]. Therefore, tracing a prediction to physically meaningful predictive signals supported by the historical record can help assess whether it provides a credible basis for issuing an early warning.

Such predictive signals can be investigated through historical climate diagnostics and analyses of model behavior. Historical analogue analyses identify past cases resembling the target anomaly, and composite analyses assess whether they share recurrent patterns of atmospheric circulation, moisture transport, or oceanic conditions[13–15]. At the model level, feature attribution methods in explainable artificial intelligence (XAI) identify the input regions and variables that contribute to a particular prediction[9,16–22]. Because attribution results can vary among methods, their reliability is commonly assessed through cross-method comparisons[23,24]. Perturbation tests examine whether changing the identified features alters the prediction as expected[25,26].

Evidence for the credibility of an individual prediction is strengthened when historical diagnostics and feature attribution converge on the same physically interpretable signals.

Dynamical–statistical bridging provides a suitable framework in which these forms of evidence can be brought together. Such approaches translate dynamical predictions of relatively predictable large-scale climate modes into regional precipitation anomalies through statistical relationships[27–30]. More recently, deep-learning models have expanded this framework by learning nonlinear relationships between multivariable circulation fields and regional precipitation[31–33]. This explicit circulation-to-precipitation mapping allows feature attribution to identify physically interpretable circulation signals underlying an individual prediction. Yet this interpretive capability has rarely been used to support physically grounded assessment when seasonal precipitation predictions are issued in real time[11,34].

Here, we use a circulation-to-precipitation bridging model to generate real-time predictions of summer 2026 precipitation over China. Predictions initialized in March, April and May consistently indicate a dry anomaly over central China. We assess the credibility of the predicted anomaly by evaluating model skill in historical analogue years and examining whether the climate conditions associated with these years are also evident in 2026. We then use feature attribution to determine whether the model highlights circulation signals consistent with these climate features. We evaluate the stability of the attributions across methods and use LRP-guided perturbation tests to assess the relevance of the identified features to the predicted anomaly. Together, these analyses establish a framework for assessing AI-based seasonal precipitation predictions through convergent historical, physical and model-centered evidence available ahead of the target season.

## *Results*

### *1. Consistent predictions point to a central China dry anomaly in summer 2026*

We generated real-time predictions of summer 2026 precipitation over China by applying the circulation-to-precipitation bridging model to dynamical circulation predictions initialized in March, April and May. All three predictions showed a broadly consistent spatial pattern, with below-normal precipitation across central China and above-normal precipitation over South China and parts of North China (Fig. 1**a–c**). The central China dry anomaly was the most prominent regional feature, with regional means of −34%, −29% and −38% in the three predictions, respectively. Although central China showed the largest anomaly magnitude, its inter-initialization range was comparatively small, whereas larger ranges occurred over parts of South and Northeast China (Fig. 1**d**). The location, spatial extent and magnitude of the central China dry anomaly therefore remained relatively stable as the initialization approached the target season, indicating the anomaly was not specific to a particular initialization month.

Cross-validation over 1993–2025 showed comparable skill between the bridging-model and dynamical multi-model ensembles (MMEs; Fig. 1e, f). The bridging-model MME yielded anomaly correlation coefficient (ACC) values of 0.140–0.161 and prediction scores (Ps) of 73.27–74.25% across the three initialization months. Relative to the dynamical MME, its largest gains occurred for the March initialization, whereas the two MMEs performed similarly in May. These results motivate a case-specific credibility assessment of the predicted dry anomaly over central China.

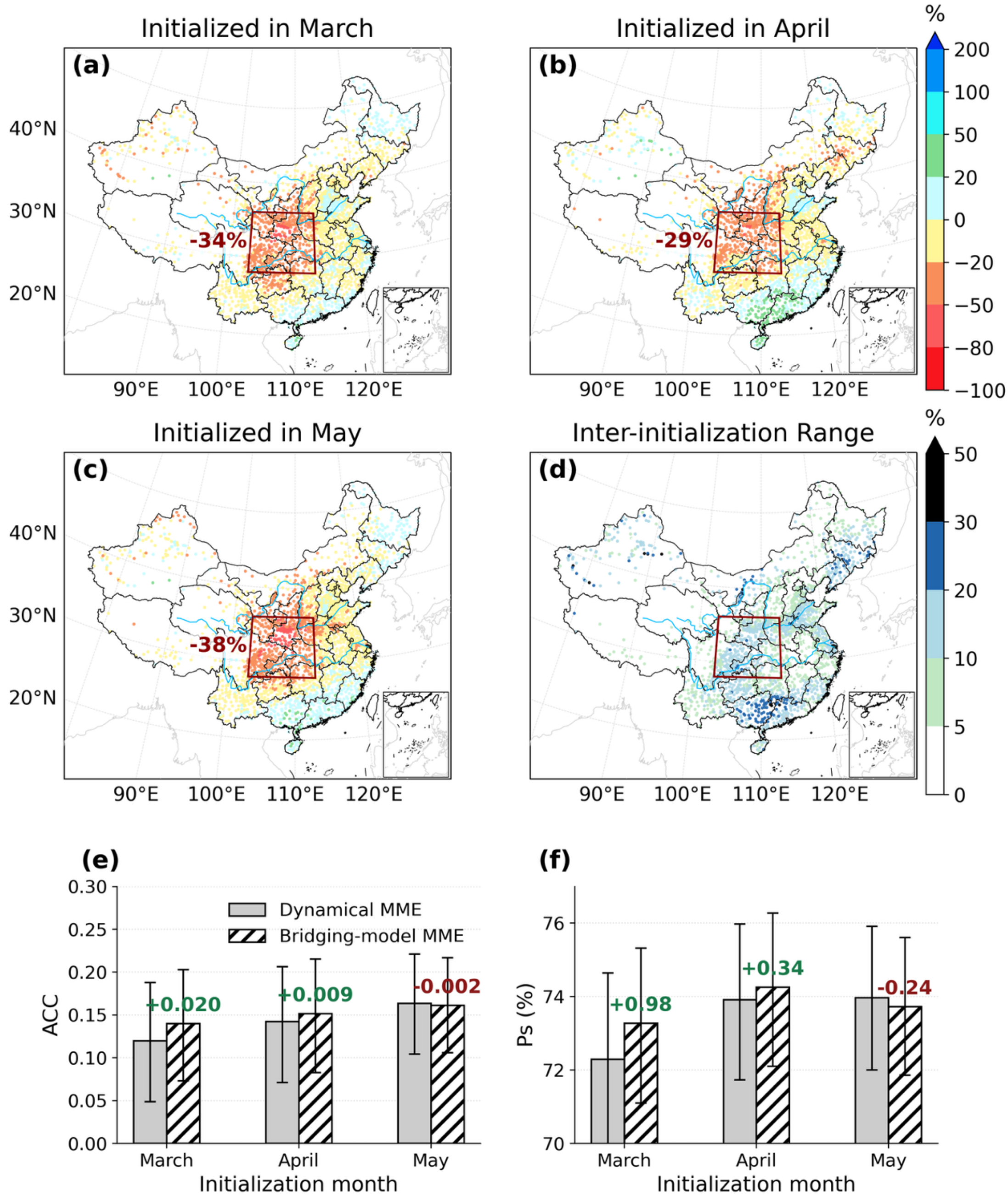

**Fig. 1: Summer 2026 precipitation predictions over China and cross-validated hindcast skill.** a–c, Precipitation anomalies (%) relative to the 1991–2020 climatology predicted by the bridging-model multi-model ensemble (MME) for June–August 2026, based on the March, April and May initializations, respectively. d, Inter-initialization range at each station, calculated as the difference between the maximum and minimum anomalies among the three predictions. Red boxes in a–d delineate central China (28°–36°N, 103°–113°E), and the values in a–c indicate the corresponding regional-mean precipitation anomalies. e, f, Historical cross-validated anomaly correlation coefficient (ACC; e) and prediction score (Ps; f) for summer precipitation predictions initialized in March, April and May during 1993–2025. Gray and hatched bars represent the raw dynamical MME and the bridging-model MME, respectively. Error bars indicate bootstrap 95% confidence intervals for the mean cross-validated scores.

Values above the bars indicate the differences between the bridging-model and raw dynamical MMEs.

## 2. *Higher prediction skill in historical dry-anomaly years*

The historical cases most closely resembling the predicted 2026 precipitation pattern were highly consistent across initializations. For each initialization, the observed summers during 1993–2025 were ranked separately using four metrics of similarity to the corresponding prediction. The four ranks were then summed for each year, and the seven years with the lowest rank sums were retained for analysis (Methods and Supplementary Table 1). Six years—1994, 1997, 2001, 2006, 2011 and 2015—were common to all three sets. The April and May predictions yielded identical sets, additionally including 2002, whereas the March prediction included 2022 instead.

We next compared cross-validated prediction skill between the selected dry-anomaly years and the remaining 26 years to examine how the model had performed in historical cases resembling the 2026 prediction (Fig. 2). Across all three initializations and both year groups, ERA5-driven outputs attained markedly higher ACC and Ps than predictions driven by dynamical-model circulation, suggesting that errors in the predicted circulation partly constrained real-time prediction skill. Despite their different overall skill levels, both input settings showed a common conditional pattern: median skill was higher in the dry-anomaly years in every comparison except ACC for the May-initialized, dynamically driven predictions, for which the dry-year median was slightly lower and the two distributions overlapped substantially. These comparisons suggest that the model may perform better for precipitation states resembling the predicted 2026 dry anomaly than its average historical skill would imply. Because the bridging-model predictions are derived entirely from atmospheric circulation, the higher skill in the dry-anomaly years suggests that a recurrent circulation pattern may underlie the model's prediction of central China drying.

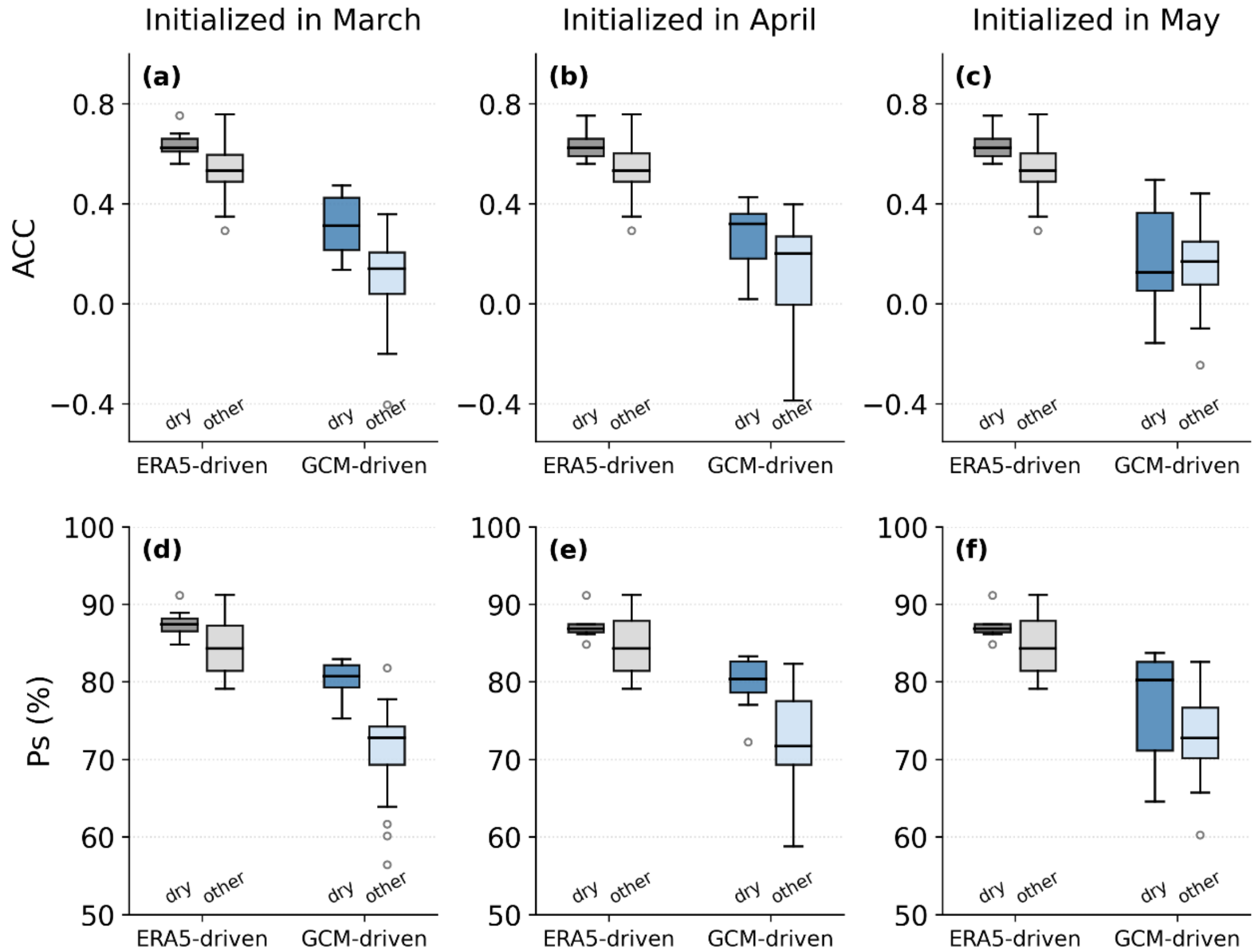

**Fig. 2: Prediction skill in analogue years for the predicted 2026 dry anomaly. a–c**, ACC of summer precipitation outputs for the March, April and May initializations during 1993–2025, grouped into the 7 dry-anomaly analogue years selected for each initialization and the remaining 26 years. **d–f**, corresponding Ps. Gray boxplots represent bridging-model outputs driven by ERA5 circulation, whereas blue boxplots represent bridging-model predictions driven by dynamical-model circulation. Within each input setting, darker and lighter shades denote analogue years and other years, respectively.

### *3. Climate conditions conducive to a dry anomaly over central China*

We composited the summer circulation across the seven analogue years selected for the May-initialized bridging-model prediction and compared them with the May-initialized dynamical MME circulation prediction for summer 2026 (Fig. 3). The March- and April-initialized predictions showed similar spatial patterns (Supplementary Figs. 1 and 2).

In the upper troposphere, the analogue composite showed a northward displacement of the East Asian subtropical westerly jet (EASWJ), with strengthened 200-hPa zonal winds north of the climatological jet axis and weakened winds to its south (Fig. 3**e**). The May-initialized prediction reproduced this meridional dipole (Fig.

3**i**). Previous observational studies have associated a poleward-displaced EASWJ with reduced summer rainfall over the Yangtze–Huaihe River Valley and a redistribution of rainfall toward South and Northeast China[35–37].

In the middle troposphere, the composite featured a pronounced positive 500-hPa geopotential-height anomaly extending from Mongolia to south of Lake Baikal (Fig. 3**f**). This type of continental high anomaly has been associated with enhanced subsidence south of Lake Baikal, anomalous low-level northerly winds over eastern China, and a weakened East Asian summer monsoon with reduced moisture transport[38]. The May-initialized prediction reproduced the positive-height anomaly but placed its center farther west (Fig. 3**j**).

At 700 hPa, both the dry-anomaly composite and the 2026 prediction featured a cyclonic circulation anomaly over the western North Pacific that extended westward into the South China Sea–South China region (Fig. 3**g, k**). Central China lay on the northwestern flank of this circulation, where anomalous northeasterly flow was stronger in the 2026 prediction than in the historical composite. Consistent with this circulation, the 850–500-hPa layer-integrated moisture-flux anomaly opposed the climatological moisture transport into central China, and the target region showed anomalous moisture-flux divergence (Fig. 3**d, h, l**). The 2026 prediction also showed anomalous moisture-flux convergence over South China, consistent with the above-normal precipitation predicted there (Figs. 3**l** and 1).

Among these circulation anomalies, the western North Pacific–South China Sea–South China region cyclonic anomaly most directly favored central China drying by weakening moisture transport into the target region, with the stronger anomalous winds associated with this circulation in the 2026 prediction suggesting that its drying influence may be stronger than in the composite of the seven analog years.

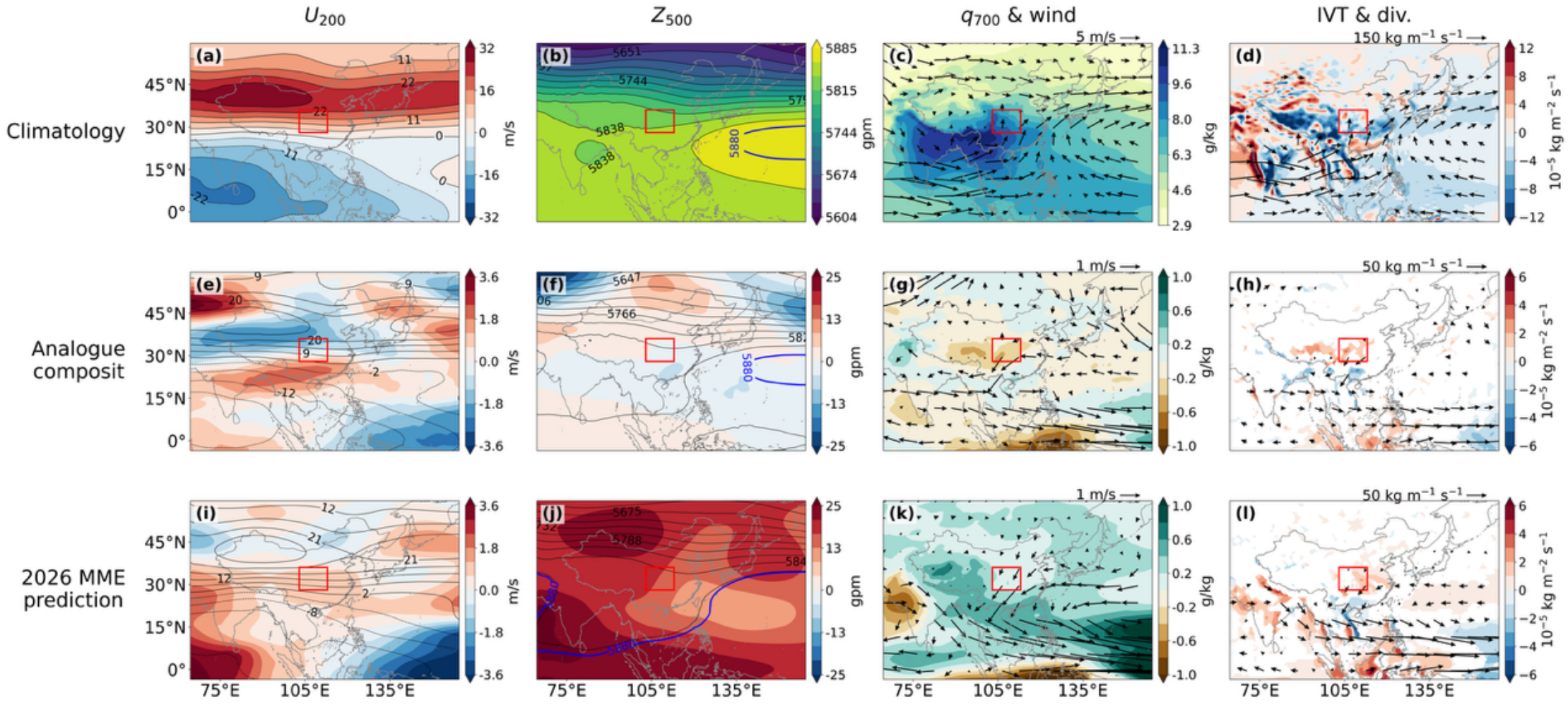

**Fig. 3: Circulation patterns associated with the predicted central China dry anomaly. a–d**, ERA5 JJA climatology for 1993–2016. **e–h**, Composite anomalies for the seven historical dry-anomaly years selected for the May initialization. **i–l**, May-initialized dynamical MME anomalies for JJA 2026. Columns show 200-hPa zonal wind ($U_{200}$; shading and contours; m $s^{-1}$), 500-hPa geopotential height ($Z_{500}$; shading and contours; gpm), 700-hPa specific humidity ($q_{700}$; shading; g $kg^{-1}$) and horizontal winds (vectors; m $s^{-1}$), and 850–500-hPa integrated vapor transport calculated from the 850-, 700- and 500-hPa levels (IVT; vectors; kg $m^{-1}$ $s^{-1}$) and its divergence (div.; shading; $10^{-5}$ kg $m^{-2}$ $s^{-1}$). Red boxes delineate the central China target region. Blue 5880-gpm contours in **b**, **f** and **j** delineate the western North Pacific subtropical high. In **d**, IVT vectors are shown only where their magnitude is ≥50 kg $m^{-1}$ $s^{-1}$. In **h**, vectors are retained where their magnitude is ≥6 kg $m^{-1}$ $s^{-1}$ and at least one horizontal component differs from zero in a two-sided one-sample t-test ($p < 0.20$); divergence shading is restricted to grid points satisfying the same significance threshold. In **l**, vectors are shown where their magnitude is ≥6 kg $m^{-1}$ $s^{-1}$, and divergence shading is retained where at least six of the seven models agree on the sign of the MME-mean divergence anomaly.

The recurrent circulation pattern was accompanied by a tendency towards warmer conditions in the central equatorial Pacific. Niño-4 SST increased markedly from the preceding winter into summer in six of the seven historical dry-anomaly years, whereas the central equatorial Pacific was already anomalously warm in the preceding winter and remained so throughout 2015 (Fig. 4). In 2026, Niño-4 rose continuously from a negative anomaly in December 2025 to above 1 °C by May 2026. Recent climate monitoring indicates that El Niño conditions have developed in the tropical Pacific (https://wmo.int/news/media-centre/el-nino-forecast-intensify-increasing-likelihood-of-extreme-weather;

https://www.cpc.ncep.noaa.gov/products/analysis_monitoring/enso_advisory/ensodisc

.shtml; https://ds.data.jma.go.jp/tcc/tcc/products/elnino/outlook.html ), and forecasts further suggest that the developing El Niño intensity could become exceptionally strong[39]. Although the strongest warming in 2026 was located farther east, the warming extended into the Niño-4 region. Previous studies found that the relationship between developing El Niño and reduced summer rainfall over central China became significant after the late 1980s, owing to the increased frequency of central-Pacific El Niño events[40]. When SST warming is centered over the central equatorial Pacific, the associated Walker circulation anomaly shifts westward, extending its descending branch to the Maritime Continent and favoring a lower-tropospheric cyclonic circulation anomaly over the western North Pacific–South China Sea[41]. The cyclonic circulation weakens northward monsoon moisture transport[40]. The pronounced Niño-4 warming in 2026 is consistent with this mechanism and the predicted reduction in summer rainfall over central China.

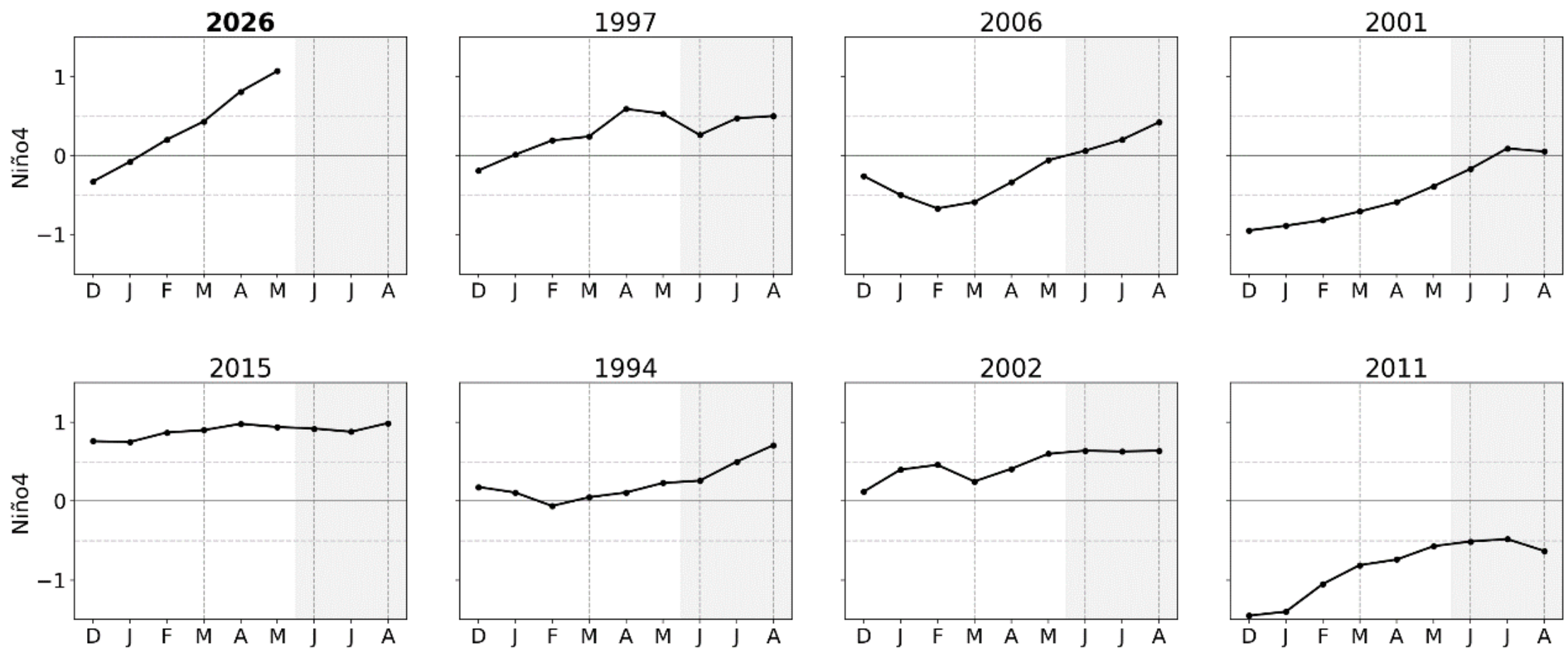

**Fig. 4: Niño-4 SST anomaly evolution in 2026 and historical analogue years.** Monthly Niño-4 SST anomalies (°C) from December of the preceding year to August of each of the seven historical dry-anomaly years selected for the May initialization. The 2026 series spans December 2025–May 2026. Gray shading marks the JJA target season.

### *4. Physically coherent prediction signals identified by LRP*

To examine whether the bridging model drew on the circulation features highlighted in the previous section, we applied LRP to the May-initialized prediction for summer 2026 (Fig. 5). We defined the attribution target as the predicted regional-mean precipitation anomaly across the central China stations. The linear mapping from the predicted principal component (PC) coefficients through empirical orthogonal function (EOF) reconstruction and regional averaging was incorporated as the final target layer for relevance propagation, with each mode weighted by its contribution to the regional anomaly. This formulation ties the resulting attribution directly to the regional prediction of interest, avoiding separate interpretation and subsequent combination of mode-specific attribution maps.

A physically coherent attribution pattern was evident in the meridional wind fields at 500, 700 and 850 hPa (Fig. 5**d, g, j, n**). At all three levels, dry-supporting relevance was concentrated on the northerly anomalies along the northwestern flank of the western North Pacific–South China Sea–South China region cyclonic circulation, with a particularly pronounced signal at 700 hPa. Together, the three meridional-wind predictors accounted for 37.5% of the area-weighted dry-supporting relevance. These northerly anomalies oppose the climatological moisture transport into central China. The predicted fields showed widespread positive specific-humidity anomalies at 500, 700 and 850 hPa, while the corresponding predictors received little dry-supporting relevance (Fig. 5**e, h, k, n**). These results indicate that the model associated the predicted rainfall deficit primarily with circulation-induced weakening of moisture transport into central China rather than with limited moisture availability. The spatial distribution of this attribution closely matched the circulation pathway identified independently from the analogue composites and the 2026 dynamical prediction.

Beyond the meridional-wind signals, dry-supporting relevance in 2-m air temperature (T2m) was concentrated over an anomalously warm region upstream of central China (Fig. 5**m, n**). This spatial correspondence indicates that the model associated the temperature signal with the predicted dry anomaly. The 500-hPa geopotential height (Z500) field also received dry-supporting relevance (Fig. 5**b, n**), concentrated mainly along strong geopotential-height gradients on the WNPSH periphery. However, the alignment of its edges with the Vision Transformer (ViT) patch grid suggests that some of the spatial structure may arise from tokenization artifacts.

Integrated Gradients (IG) and Guided Integrated Gradients (Guided IG) broadly reproduced the May-initialized wind-field attribution patterns (Supplementary Figs. 3 and 4). Cross-method agreement in predictor rankings remained high across all three initializations (Supplementary Table 2).

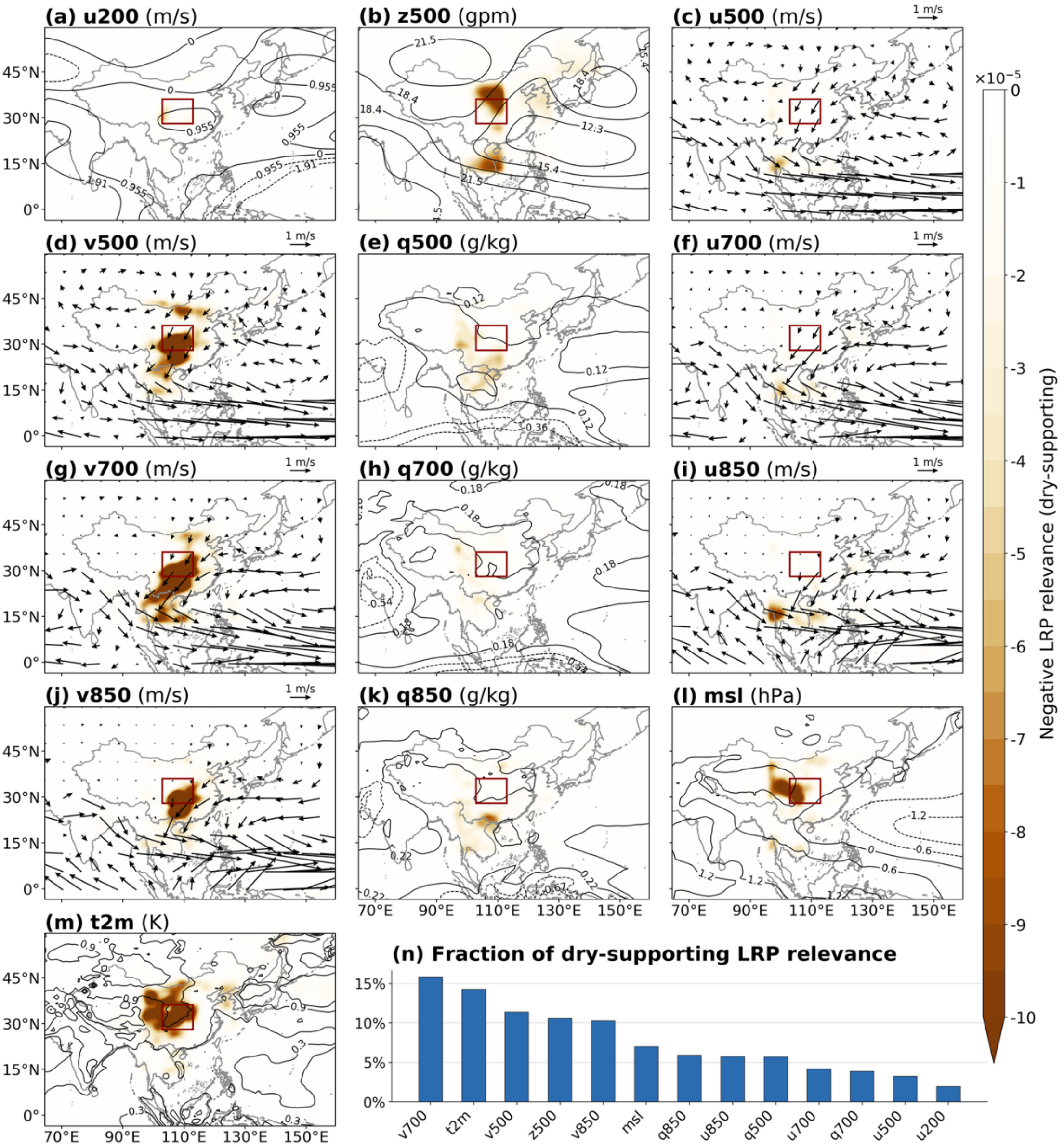

**Fig. 5: LRP-identified circulation signals supporting the predicted summer 2026 dry anomaly over central China. a–m**, Spatial distributions of dry-supporting negative LRP relevance for the 13 circulation predictors, calculated for the regional-mean precipitation anomaly over central China from the May initialization. High negative relevance indicates a stronger contribution toward the predicted dry anomaly; only negative, dry-supporting

relevance is shown, and the fields are smoothed using a Gaussian filter with σ=1 grid cell. Contours denote the corresponding predicted anomalies for scalar fields, and vectors show horizontal wind anomalies at 500, 700 and 850 hPa in the paired u and v panels. The red box marks the target region. **n**, Relative contribution of each predictor, calculated from the spatially integrated negative relevance. Here, u, v, q and z denote zonal wind, meridional wind, specific humidity and geopotential height, with pressure levels indicated by the subscripts; msl and t2m denote mean sea-level pressure and 2-m air temperature.

### *5. Perturbation tests support the faithfulness of LRP explanations*

To test whether the circulation signals highlighted by LRP materially influenced the predicted dry anomaly, we conducted LRP-guided Removal and Retention tests for the May initialization (Fig. 6). Perturbation responses were evaluated through the regional LRP target, $\Phi_R$, which is equivalent to the mean standardized precipitation anomaly across central China stations; positive values indicate wet anomalies and negative values indicate dry anomalies. For 2026 and its seven analog years, case-specific masks were constructed from the variable–grid-cell features with the top 5% of dry-supporting relevance and their neighboring grid cells (Fig. 6**a**). The masked features were zeroed out in the Removal test and retained exclusively in the Retention test. Each test was repeated with 100 random masks perturbing the same fraction of the input. The prespecified one-sided comparisons tested whether Removal yielded a higher $\Phi_R$ and Retention a lower $\Phi_R$ than the corresponding random-mask controls.

LRP-guided Removal reversed the sign of $\Phi_R$ from negative to positive for 2026 and all seven analog years, whereas random Removal left $\Phi_R$ close to the baselines (Fig. 6**b**). Conversely, LRP-guided Retention preserved negative $\Phi_R$ and produced more negative values than the corresponding baselines in every case, while random Retention yielded values clustered near zero (Fig. 6**c**). None of the 100 random perturbations produced a response more extreme than the LRP-guided result in the prespecified direction for any of the eight cases (one-sided empirical tests, all $P = 0.0099$). The disappearance of the dry signal after Removal and its persistence under Retention indicate that the circulation signals supporting the predicted dry anomaly were concentrated in the input features highlighted by LRP, providing evidence that the attribution faithfully reflected the model's dependence on these inputs.

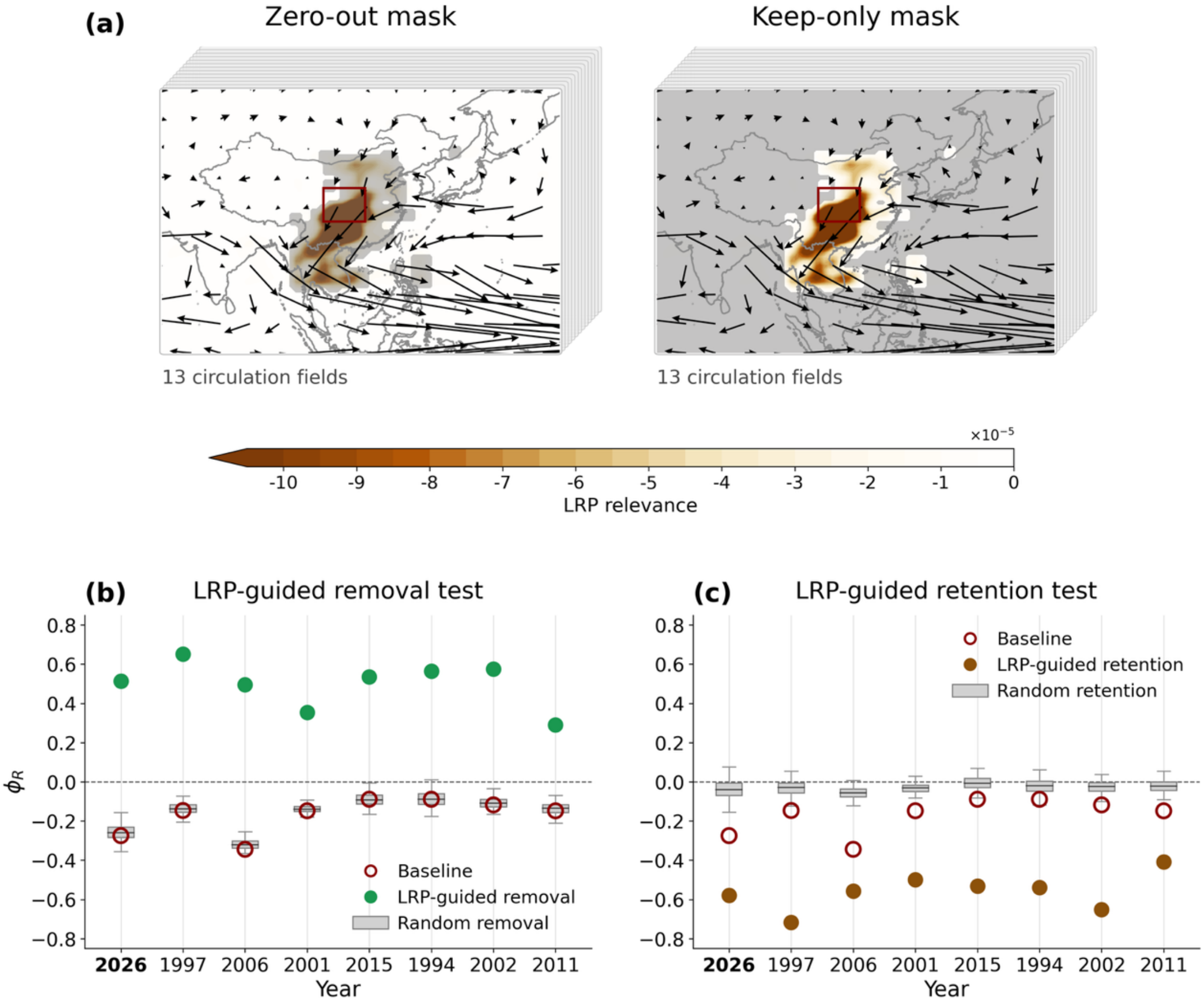

**Fig. 6: LRP-guided perturbation tests of central China dry-anomaly predictions. a**, Schematic of the zero-out and keep-only masks applied across the 13 circulation input fields for the May initialization. The LRP-guided masks comprise the variable–grid-cell features with the top 5% of dry-supporting relevance and their neighboring grid cells; the red box delineates central China. **b**, **c**, $\Phi_R$ for 2026 and its seven analog years under the Removal (**b**) and Retention (**c**) tests. Open red circles show the baselines, colored dots the LRP-guided results, and gray boxplots the distributions from 100 random-mask controls perturbing the same fraction of the input. $\Phi_R$ denotes the regional LRP target.

***Discussion***

This study provides a prospective, case-specific assessment of the predicted summer 2026 dry anomaly over central China using only information available at each prediction start time. Historical evaluation showed generally higher skill in the selected analogue years than in other years. The predicted circulation also reproduced features associated with suppressed precipitation over central China, including a lower-tropospheric cyclonic anomaly over the western North Pacific–South China Sea–South China region, northerly anomalies on its northwestern flank, weakened monsoon moisture transport, and moisture-flux divergence over central China. LRP indicated that the model relied strongly on these northerly anomalies. In perturbation tests, removing the LRP-selected features reversed the predicted anomaly from dry to wet, whereas retaining these features alone intensified the dry anomaly, supporting the faithfulness of the LRP attribution.

In this study, we defined the LRP target as the predicted regional-mean precipitation anomaly over central China. The network outputs 512 PC coefficients, and the regional anomaly is reconstructed from all 512 associated EOF modes. Attributing each PC separately would not directly explain why central China was predicted to be dry and would require combining many mode-specific maps. Thus, we included the linear reconstruction from all 512 PCs to the regional mean in the relevance-propagation pathway. The resulting map integrates all modes into a single region-specific attribution that can be compared directly with the circulation patterns. This formulation may also be useful for regional interpretation of other spatial prediction models with reduced-order outputs.

LRP attributes the model output to input features but does not establish causal relationships in the climate system. Evidence was strongest for the mid-to-lower-tropospheric northerly signal, which agreed with climate diagnostics and was broadly reproduced by IG and Guided IG. Secondary signals require caution. In particular, T2m is a rapidly responding component of the coupled land–atmosphere system. Reduced precipitation and cloud cover, together with diminished soil moisture and evaporative cooling, can increase near-surface temperature, producing a coupled warm–dry surface state. The bridging model can exploit this covariation, and the dry-supporting T2m relevance over northwestern China may indicate that such a coupled warm–dry state contributed predictive information. It does not demonstrate that warming over northwestern China directly caused the predicted precipitation deficit over central China. The patch-aligned Z500 relevance also suggests that its detailed spatial pattern may partly reflect ViT tokenization. The perturbation tests confirmed the joint importance of the selected features to the model prediction, but did not establish them as causal drivers in the climate system.

The framework examines whether similar historical cases were more predictable, whether the predicted circulation supports a plausible physical pathway, and whether the model relies on features associated with that pathway. For summer 2026, all three analyses provided supportive evidence. The value of this framework lies in making the evidential basis of a prediction explicit before observations become available, rather than guaranteeing that the prediction will verify. The large-scale conditions captured by the framework may modulate subseasonal rainfall variability and the likelihood of extreme events, but the model does not explicitly resolve the occurrence, timing or intensity of individual events. The predicted dry anomaly therefore represents a seasonal tendency toward drier conditions and does not rule out the possibility of heavy rainfall that could substantially affect the observed summer precipitation total. More broadly, it provides physically grounded explanations for seasonal predictions of potentially high-impact climate anomalies and supports the transparent and evidence-based use of AI in pre-season climate-risk assessment.

## *Methods*

### *1. Data*

Monthly station precipitation observations for 1961–2025 were calculated from the *China Precipitation Daily Dataset (V3.0)*[42], provided by the National Meteorological Information Centre of the China Meteorological Administration. These observations served as the predictand for transfer learning, the reference for cross-validation evaluation, and the basis for identifying historical analogue years. Monthly Niño-4 index were obtained from the NOAA Physical Sciences Laboratory climate-index archive. The index is produced by the NOAA Climate Prediction Center from ERSSTv5[43] and represents the area-mean SST anomaly over 5° S–5° N, 160° E–150° W.

Monthly mean atmospheric fields for 1961–2025 were obtained from ERA5[44]. The fields were interpolated onto a 1° × 1° grid covering 64.5°–159.5° E and 3.5° S–59.5° N. We used 13 variables: zonal wind at 200 hPa; geopotential height, zonal and meridional winds, and specific humidity at 500 hPa; zonal and meridional winds and specific humidity at 700 and 850 hPa; mean sea-level pressure; and 2-m air temperature. ERA5 circulation fields served as model inputs during transfer learning and retrospective evaluation. They were also used to diagnose circulation and moisture transport in the analogue years.

Seasonal prediction data were obtained from the monthly pressure-level and single-level datasets in the Copernicus Climate Change Service (C3S) Climate Data Store[45,46]. The hindcast for 1993–2025 comprised 27 forecast-system versions from eight forecasting centers; the systems and their ensemble sizes are listed in Table 1, with ensemble sizes given in parentheses. The same 13 atmospheric variables and the corresponding precipitation formed paired inputs and targets for pretraining. Because the operational system composition changed during spring 2026, the March initialization used seven systems (BOM-2, CMCC-4, DWD-22, ECCC-5, ECMWF-51, Météo-France-9 and UKMO-605), whereas the April and May initializations used eight (BOM-2, CMCC-4, DWD-22, ECCC-5, ECMWF-51, JMA-4, Météo-France-9 and UKMO-610). Each system was processed separately, and the resulting precipitation predictions were averaged with equal system weights to form the MME prediction. All monthly data were converted to overlapping three-month means, yielding 12 seasonal samples per year.

| forecasting center | model |
|---|---|
| ECMWF | SEAS5 (25), SEAS5.1 (25) |
| MF | Météo-France System 6 (25), 7 (25), 8 (25), 9 (31) |
| UKMO | GloSea5-GC2 12 (28), 13 (28), GloSea5-GC2-LI 14 (28), 15 (28), GloSea6 600 (28), 601 (28), 602 (28), 603 (28), 604 (28) |
| CMCC | CMCC-SPS 3 (40), 3.5 (40), 4 (30) |
| DWD | GCFS 2.0 (30), 2.1 (30), 2.2(30) |
| ECCC | GEM5-NEMO (10), CanESM5.1p1bc (20), GEM5.2-NEMO (20) |
| JMA | JMA-CPS 2 (10), 3 (10) |
| BOM | ACCESS-S2 (27) |

**Table 1. Forecasting centers and C3S seasonal prediction systems used for model pretraining.**

### *2. Bridging-model development and evaluation*

The circulation-to-precipitation bridging model maps 13 seasonal circulation-anomaly fields to station precipitation anomalies (Fig. 7). The model is an updated version of the circulation-to-precipitation bridging model developed by Jin et al.[32], with a ViT architecture replacing the original convolutional backbone. The multichannel fields were divided into patches and transformed into spatial tokens through convolutional patch embedding and bottleneck projection. The target three-month season was embedded as a classification (CLS) token, which was prepended to the spatial-token sequence and used to condition feature extraction through adaptive layer normalization. After positional embeddings were added, the sequence was processed by six ViT blocks. The updated CLS token was passed through root mean square normalization, dropout and two fully connected layers to predict the PC coefficients of the leading 512 EOF modes. Station precipitation anomalies were then reconstructed from the predicted coefficients and the EOF basis.

Because each three-month window yields only one seasonal sample, the observation dataset is relatively small. The model was therefore pretrained using paired circulation and precipitation anomalies from dynamical models. During transfer learning, the final two linear layers were fitted by ridge regression using ERA5 circulation anomalies and observed precipitation anomalies.

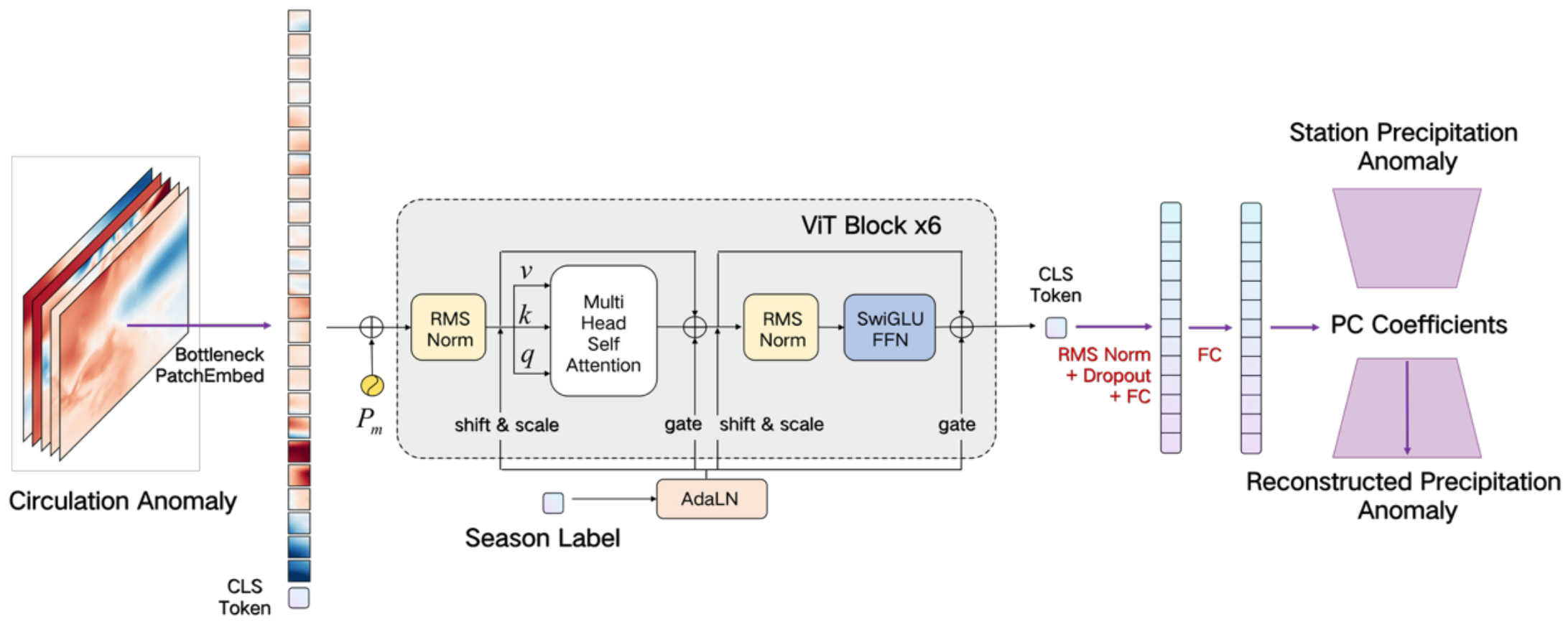

**Fig. 7: Architecture of the circulation-to-precipitation bridging model.** Spatial information is integrated within each ViT block via Multi-Head Self-Attention (MHA), while target-season information from the CLS token modulates the main branch of the neural network through shifting, scaling, and gating. Purple trapezoids indicate the EOF projection used to derive the PC targets and reconstruct station precipitation anomalies. $P_m$, positional embedding; CLS, classification; PC, principal component.

Cross-validation for 1993–2025 used six contiguous year blocks. In each fold, one block was held out for testing, the preceding block in cyclic order for validation, and the remaining four for training. The same splits were applied during pretraining and transfer learning. Predictions for the six test blocks formed the complete cross-validated record.

Prediction performance was evaluated using the spatial ACC and $P_s$. For $N$ stations, ACC was calculated as

$$ACC = \frac{\sum_{i=1}^{N}(p_i - \bar{p})(o_i - \bar{o})}{\sqrt{\sum_{i=1}^{N}(p_i - \bar{p})^2 \sum_{i=1}^{N}(o_i - \bar{o})^2}}$$

where $p_i$ and $o_i$ are the predicted and observed precipitation anomaly percentages at station $i$, and the overbars denote spatial means. Following the National Climate Center specification[47], the $P_s$ is an empirical operational metric that combines anomaly-grade accuracy with grade-dependent weighting of precipitation anomalies:

$$P_s = \frac{2 \times N_0 + 2 \times N_1 + 4 \times N_2}{N + N_0 + 2 \times N_1 + 4 \times N_2 + \mathrm{M}} \times 100\%$$

Here, $N$ is the total number of stations; $N_0$ denotes correct anomaly-sign predictions, $N_1$ and $N_2$ denote correctly predicted level-1 and level-2 anomalies, and $M$ denotes the number of stations at which the observed precipitation anomaly was at least 100%, but the predicted anomaly failed to reach the second-level positive-anomaly category. Level-2 and level-1 anomalies correspond to $20\% \leq| \Delta R |< 50\%$ and $| \Delta R |\geq 50\%$.

### *3. Identification of historical analogue years*

For each initialization, observed JJA precipitation anomalies during 1993–2025 were compared with the corresponding 2026 prediction using four metrics: all-station ACC ($\mathrm{ACC_{all}}$), Central China ACC ($\mathrm{ACC_{CC}}$), dry coverage (DC), and dry-intensity distance (DID). Let $P_y$ denote the observed precipitation anomaly percentage at Central China stations in year $y$, $\hat{P}$ the corresponding prediction for 2026. DC and DID were calculated as

$$\mathrm{DC_y} = \frac{N_y^-}{N_{CC}} \times 100\%, \ \mathrm{DID_y} = \left|P_y - \hat{P}\right|$$

where $N_y^-$ is the number of Central China stations with negative anomalies and $N_{CC}$ is the total number of stations in the region. Years were ranked separately for each metric, with rank 1 assigned to the closest match: larger values for the two ACCs and DC, but smaller values for DID. The overall rank sum was

$$S_y = r_y(\mathrm{ACC_{all}}) + r_y(\mathrm{ACC_{CC}}) + r_y(\mathrm{DC}) + r_y(\mathrm{DID})$$

The seven years with the smallest $S_y$ were retained as historical analogues, closely corresponding to the top 20% of ranked 33 years. Ties were resolved by smaller DID, followed by higher Central China and all-station ACCs.

### *4. Layer-wise relevance propagation and perturbation tests*

LRP attributes a scalar neural-network output to its input features by redistributing relevance backward through the network using layer-specific rules[48]. This redistribution approximately conserves total relevance between adjacent layers, yielding signed scores that indicate features supporting or opposing the selected output. In climate science, LRP has been used to interpret the spatial patterns underlying neural-network predictions, identify indicators of externally forced change, and diagnose climate states associated with enhanced predictability[22,49,50,9].

Here, we used LRP to trace the circulation signals contributing to the predicted central China dry anomaly. To accommodate the ViT architecture, we adopted the CP-LRP relevance-propagation scheme[51]. Within each self-attention layer, attention weights were held fixed and relevance was propagated through the value path, without redistribution through the query–key scores or softmax operation. Normalization and positional encoding were treated as pass-through operations. We used a γ-rule for the convolutional patch-embedding layers and an ε-rule for linear layers. LRP requires a scalar explanation target, whereas the bridging model predicts $K = 512$ precipitation PC coefficients. We therefore defined the target as the predicted mean standardized precipitation anomaly over the central China region $R$:

$$\Phi_R(x) = \frac{1}{N_R}\sum_{s\in R}\hat{P}_s(x) = \sum_{k=1}^{K}\omega_{R,k}\hat{y}_k(\mathrm{x})$$

$$\omega_{R,k} = \frac{1}{N_R}\sum_{s\in R}EOF_{s,k}$$

where $N_R$ is the number of stations within $R$, $x$ denotes the circulation input, $\hat{y}_k$ is the predicted coefficient of the $k$th EOF mode, and $\omega_{R,k}$ is the mean loading of that mode across the $N_R$ stations in $R$. This reconstruction was implemented as a fixed linear layer, allowing relevance to propagate from the regional target through the PC outputs to the circulation inputs. For visualization, only negative relevance values were retained, because they contributed toward a more negative $\Phi_R$. The relevance field for each predictor was smoothed independently using a two-dimensional Gaussian filter with $\sigma = 1\ grid$ cell. Predictor contributions were calculated by cosine-latitude-weighted integration of the dry-supporting relevance and normalized across the 13 predictors.

LRP-guided perturbation tests were conducted for the May-initialized 2026 prediction and the seven analogue years. Dry-supporting relevance values were pooled across all grid cells and all 13 predictors, and the strongest 5% were selected using a single global threshold. The resulting mask was expanded by one grid cell in each spatial direction. In the Removal test, the selected input elements were set to zero; in the Retention test, only the selected elements were retained. The increase in $\Phi_R$ after Removal and the retained $\Phi_R$ in the Retention test were compared with results from $B = 100$ random masks containing the same number of input elements as the expanded LRP mask. One-sided empirical $P$ values were calculated as

$$P = (r + 1)/(B + 1)$$

For the Removal test, $r$ was the number of random masks producing an equal or larger increase in $\Phi_R$ than the LRP-guided mask; for the Retention test, it was the number producing an equal or lower retained $\Phi_R$. The smallest attainable $P$ value was therefore 0.0099.

### *5. Integrated Gradients and Guided Integrated Gradients*

Like LRP, IG and Guided IG produce signed feature-level attributions for a specified scalar output. They were therefore used to assess whether the main dry-supporting signals identified by LRP were reproduced by methods based on different attribution principles. IG integrates gradients along a straight path from a reference state to the input[52]. Guided IG follows an adaptive path designed to reduce the accumulation of noisy gradients[53]. Both methods were applied to the frozen ViT using the regional attribution target $\Phi_R$. Because the inputs were standardized anomalies, a zero field represented the climatology of all predictors. This field served as the common baseline for evaluating how input anomalies changed the regional prediction. The same baseline and attribution procedures were used for predictions initialized in March, April and May. Negative values indicate dry-supporting contributions. The IG and Guided IG attribution fields in Supplementary Figs. 3 and 4 were smoothed for visualization in the same manner as the LRP fields. Cross-method agreement in predictor rankings and attribution patterns is shown in Supplementary Table 2.

# Interpretable AI predicts a 2026 summer dry anomaly in central China

Anran WANG [1†], Wen SHI [1†], Yong LUO [1*], Jianbin HUANG [2,3], Lijuan CHEN [4,5], Junhu ZHAO [5], Weixin JIN [6], and Huihui YUAN [1]

1 *Department of Earth System Science, Tsinghua University, Beijing* 100084*, China*

2 *Beijing Yanshan Earth Critical Zone National Research Station, University of Chinese Academy of Sciences, Beijing* 101408*, China*

3 *College of Resources and Environment, University of Chinese Academy of Sciences, Beijing* 101408*, China*

4 *State Key Laboratory of Climate System Prediction and Risk Management, National Climate Centre, China Meteorological Administration, Beijing* 100081*, China*

5 *China Meteorological Administration Key Laboratory for Climate Prediction Studies, National Climate Centre, China Meteorological Administration, Beijing* 100081*, China*

6 *Microsoft, Beijing* 100080*, China*

***Contents of this file***

[†] These authors contributed equally: Anran Wang and Wen Shi.

[*] Correspondence to: Yong Luo (yongluo@tsinghua.edu.cn)

**Supplementary Table 1. Historical analogue years selected for the summer 2026 predictions initialized in March, April and May.** For each initialization, 33 summers during 1993–2025 were ranked by all-station ACC ($ACC_{all}$), central China ACC ($ACC_{CC}$), dry coverage (DC) and dry-intensity distance (DID). The seven years with the smallest sum of metric ranks ($S_y$) were retained. DC is reported as a percentage and DID in percentage points. Six years were common to all three selections; April and May yielded identical sets, whereas March replaced 2002 with 2022.

| **Initialization** | **Year** | **$ACC_{all}$** | **$ACC_{CC}$** | **DC (%)** | **DID** | **$S_y$** | **Final rank** |
|---|---|---|---|---|---|---|---|
| March | 1997 | 0.47 | 0.86 | 93 | 1.5 | 4 | 1 |
| | 2006 | 0.38 | 0.57 | 81 | 10.3 | 9 | 2 |
| | 2001 | 0.30 | 0.59 | 79 | 16.0 | 12 | 3 |
| | 2015 | 0.18 | 0.51 | 77 | 19.4 | 21 | 4 |
| | 2022 | 0.15 | 0.42 | 74 | 15.5 | 25 | 5 |
| | 2011 | 0.20 | 0.37 | 75 | 19.0 | 25 | 6 |
| | 1994 | 0.11 | 0.54 | 77 | 19.7 | 26 | 7 |
| April | 1997 | 0.51 | 0.88 | 93 | 6.4 | 5 | 1 |
| | 2006 | 0.35 | 0.49 | 81 | 5.4 | 12 | 2 |
| | 2001 | 0.38 | 0.58 | 79 | 11.1 | 12 | 3 |
| | 2015 | 0.22 | 0.59 | 77 | 14.5 | 17 | 4 |
| | 1994 | 0.18 | 0.50 | 77 | 14.8 | 23 | 5 |
| | 2002 | 0.27 | 0.50 | 69 | 18.4 | 24 | 6 |
| | 2011 | 0.09 | 0.35 | 75 | 14.1 | 30 | 7 |
| May | 1997 | 0.57 | 0.87 | 93 | 3.2 | 4 | 1 |
| | 2006 | 0.38 | 0.55 | 81 | 15.0 | 9 | 2 |
| | 2001 | 0.34 | 0.60 | 79 | 20.7 | 12 | 3 |
| | 2015 | 0.23 | 0.54 | 77 | 24.1 | 20 | 4 |
| | 1994 | 0.16 | 0.54 | 77 | 24.4 | 24 | 5 |
| | 2002 | 0.26 | 0.45 | 69 | 28.0 | 26 | 6 |
| | 2011 | 0.16 | 0.40 | 75 | 23.7 | 27 | 7 |

**Supplementary Fig. 1. Circulation patterns associated with the March-initialized prediction.** Same as Fig. 3, but for the seven historical analogue years selected for the March initialization and the March-initialized dynamical MME prediction for JJA 2026. The analogue composite and 2026 prediction both show a western North Pacific–South China Sea–South China cyclonic circulation anomaly and anomalous moisture-flux divergence over central China.

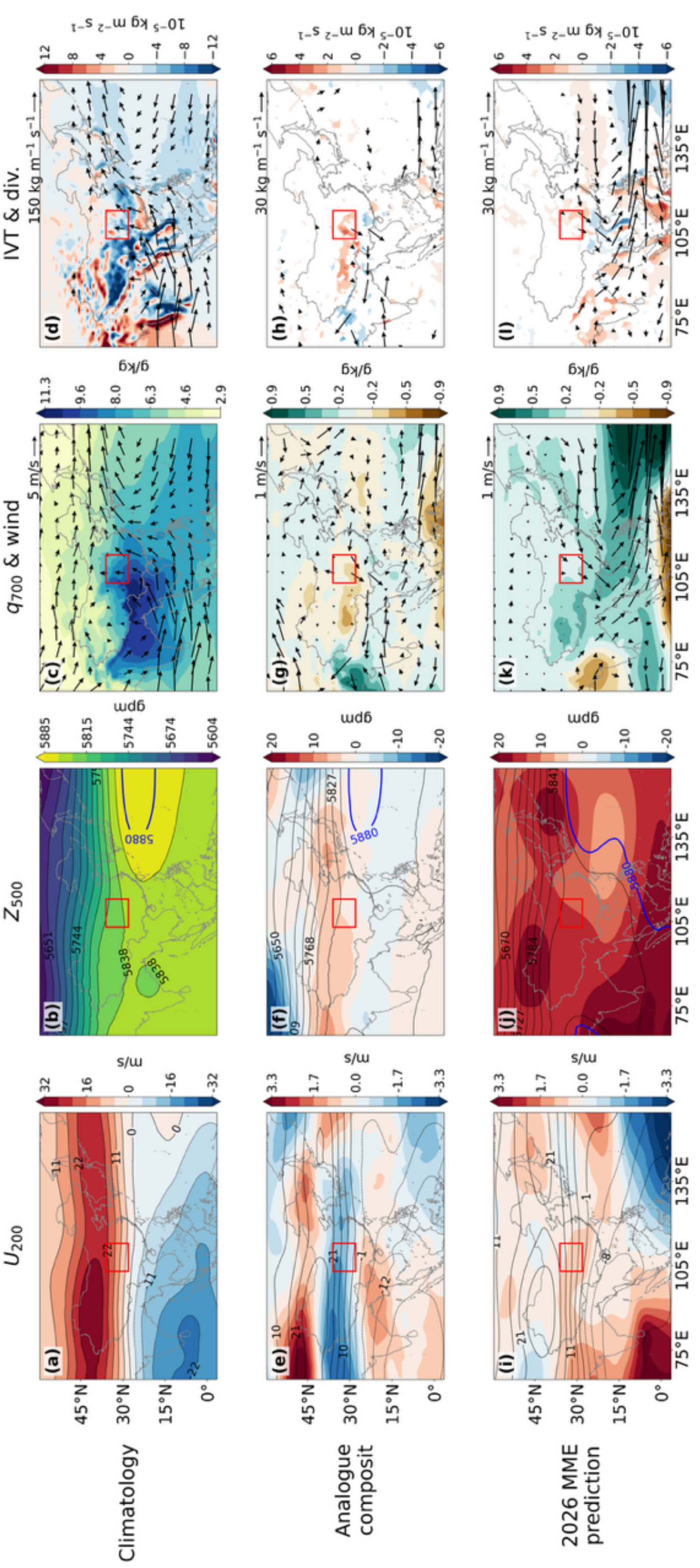

**Supplementary Fig. 2. Circulation patterns associated with the April-initialized prediction.** Same as Supplementary Fig. 1, but for the seven historical analogue years selected for the April initialization and the April-initialized dynamical MME prediction for JJA 2026.

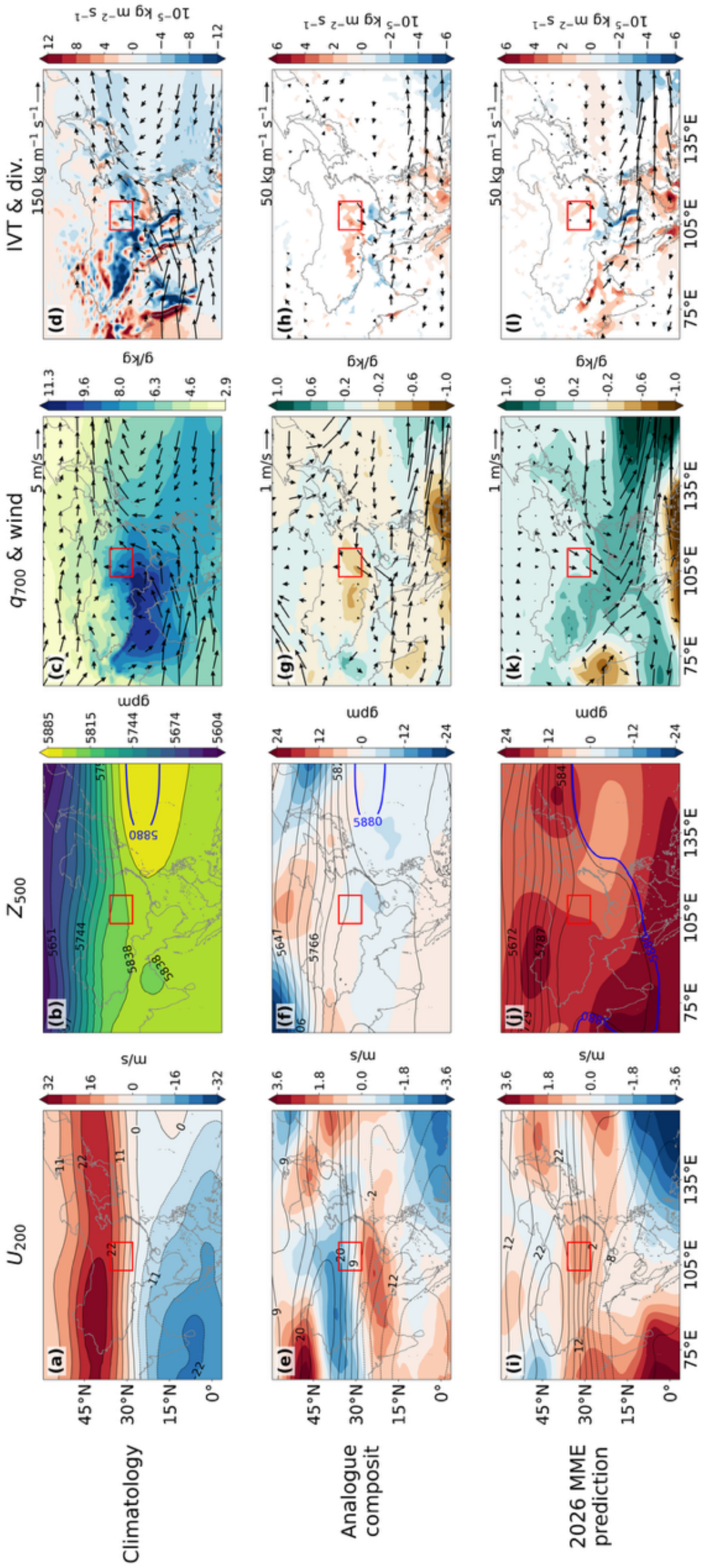

**Supplementary Fig. 3. Integrated Gradients attribution for the May-initialized prediction.** Same as Fig. 5, but showing dry-supporting negative attributions from Integrated Gradients (IG). IG broadly reproduces the dry-supporting meridional-wind signals at 500, 700 and 850 hPa identified by LRP, while assigning the largest predictor contribution to T2m.

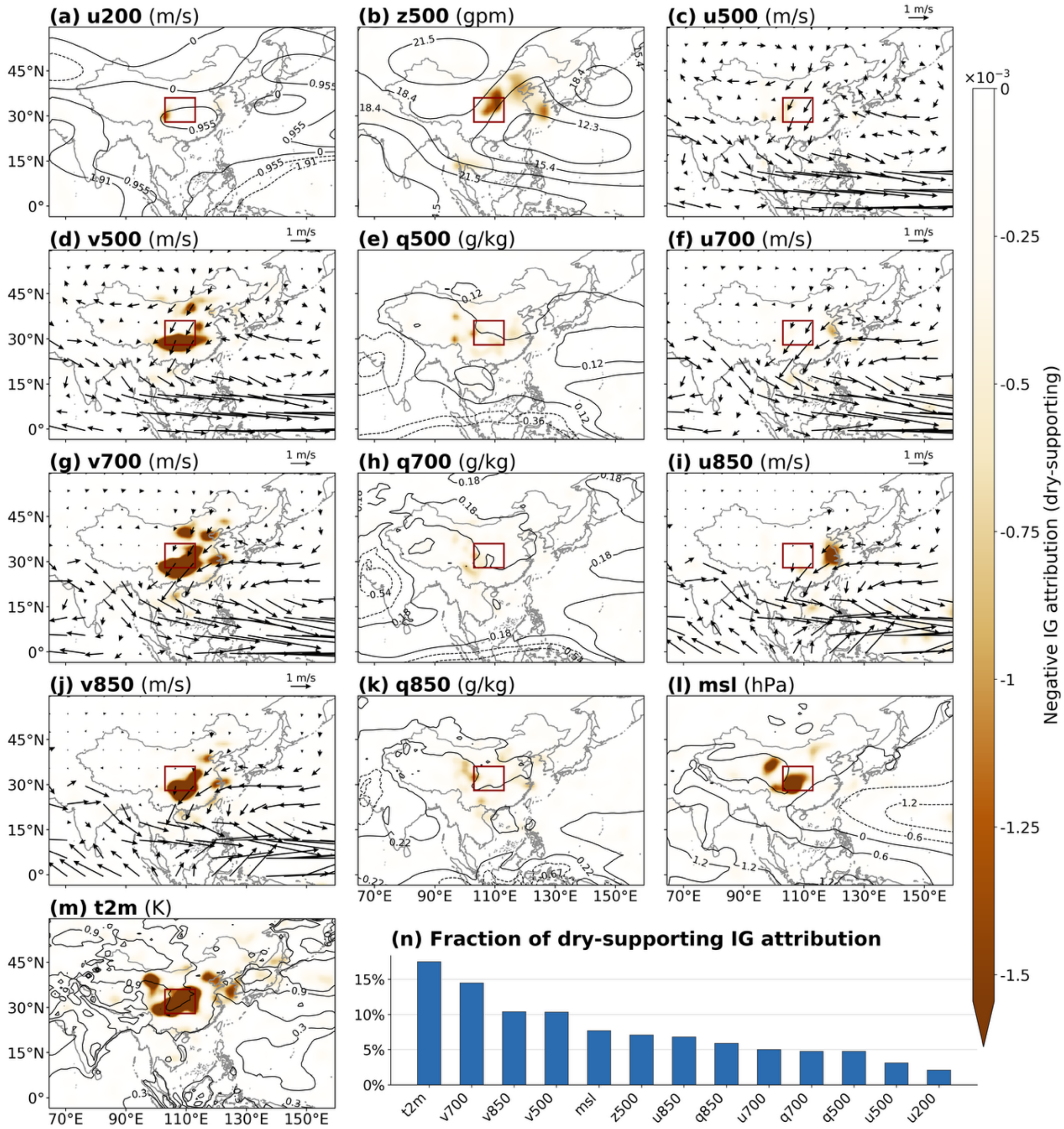

**Supplementary Fig. 4. Guided Integrated Gradients attribution for the May-initialized prediction.** Same as Supplementary Fig. 3, but using Guided Integrated Gradients (Guided IG). The dry-supporting meridional-wind signals at 500, 700 and 850 hPa are again evident, while T2m has the largest predictor contribution.

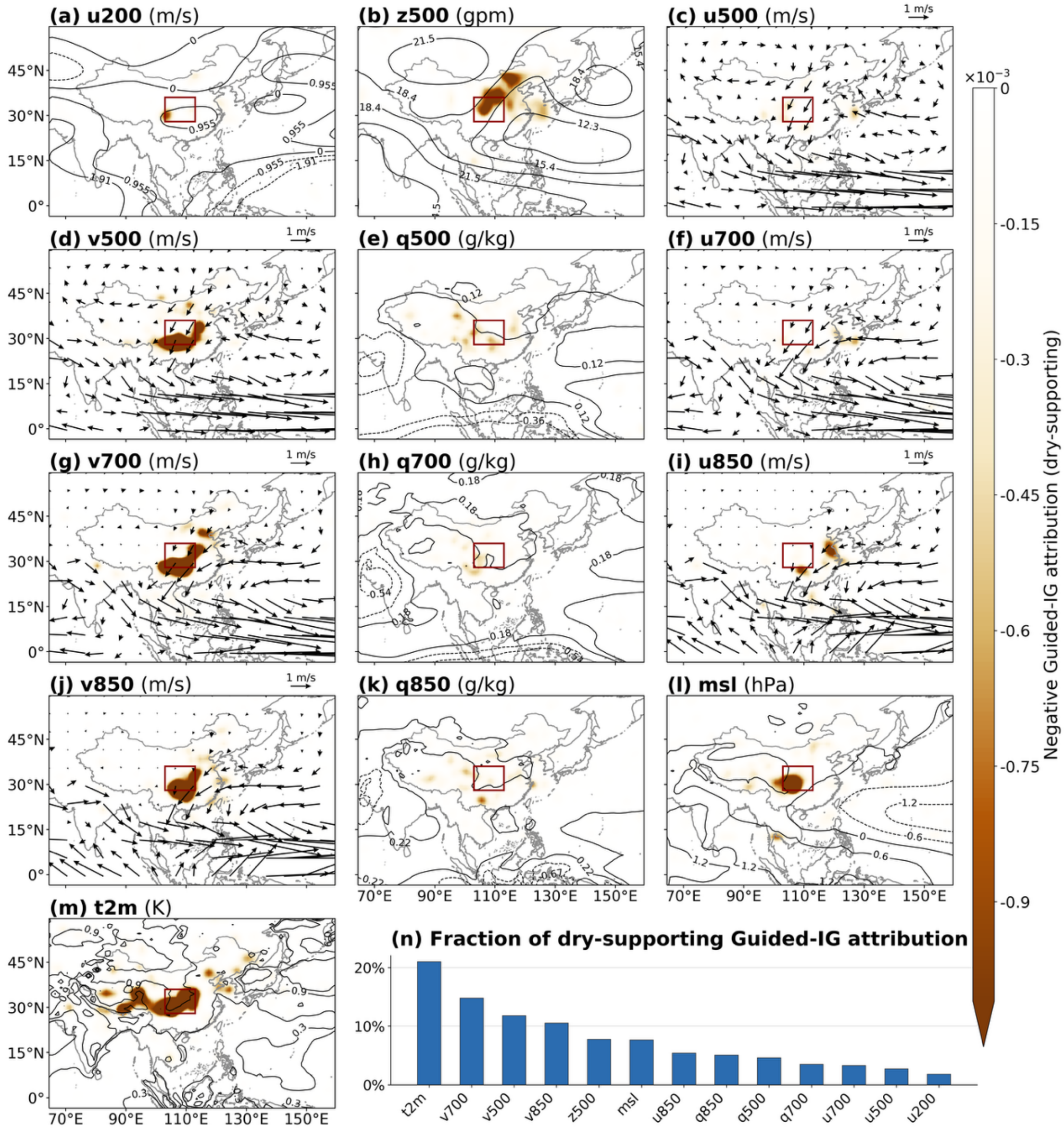

**Supplementary Table 2. Cross-method agreement in dry-supporting attributions.** Spearman rank correlation assesses agreement in the ranking of area-weighted dry-supporting contributions across the 13 predictors. Area-weighted full-tensor cosine similarity measures agreement in the unsmoothed attribution patterns across all predictors and grid cells. For both metrics, values closer to 1 indicate stronger agreement. The high rank correlations (0.88–0.98) indicate that the overall predictor hierarchy was consistent across methods. Full-tensor similarity was lower between LRP and the two gradient-based methods, indicating greater method dependence in the detailed spatial patterns. Agreement with LRP was strongest for the May initialization examined in the main text. LRP, layer-wise relevance propagation; IG, Integrated Gradients; Guided IG, Guided Integrated Gradients.

| Initialization | Method pair | Spearman rank correlation | Area-weighted full-tensor cosine similarity |
|---|---|---|---|
| March | LRP vs IG | 0.91 | 0.47 |
| | LRP vs Guided IG | 0.88 | 0.40 |
| | IG vs Guided IG | 0.95 | 0.92 |
| April | LRP vs IG | 0.92 | 0.48 |
| | LRP vs Guided IG | 0.90 | 0.43 |
| | IG vs Guided IG | 0.96 | 0.92 |
| May | LRP vs IG | **0.95** | **0.64** |
| | LRP vs Guided IG | 0.98 | 0.56 |
| | IG vs Guided IG | 0.97 | 0.90 |